\documentclass[conference]{IEEEtran}
\IEEEoverridecommandlockouts
\usepackage{amsmath,amssymb,amsfonts}
\usepackage[numbers]{natbib}
\usepackage{graphicx}
\usepackage[dvipsnames]{xcolor}

\usepackage{url}
\usepackage[acronym]{glossaries}
\usepackage{listings}
\usepackage[skip=\baselineskip]{caption}
\usepackage{booktabs}

\usepackage{xspace}
\usepackage{balance}
\usepackage{enumitem}

\lstdefinelanguage{pseudo}{
	basicstyle=\ttfamily\small,
	morekeywords={function,if,return,nil,for,from,to,mxv,apply},
	keywordstyle=\color{blue},
	breaklines=true,
	mathescape=true,
	tabsize=2,
	showstringspaces=false,
	numbers=left,
	xleftmargin=2em,
	stepnumber=1,
	numberstyle=\ttfamily\footnotesize,
	float=t,
	sensitive=false, % keywords are not case-sensitive
	morecomment=[l]{//}, % l is for line comment
	morecomment=[s]{/*}{*/}, % s is for start and end delimiter
	morestring=[b]" % defines that strings are enclosed in double quotes
}

\lstdefinelanguage{my_cpp}{
	basicstyle=\ttfamily\small,
	morekeywords={if,return,for,auto,double,void,const,size_t},
	keywordstyle=\color{blue},
	keywords=[2]{Vector,Matrix},
	keywordstyle=[2]\color{BrickRed},
	keywords=[3]{mxv,eWiseLambda,descriptors,structural,Ring},
	keywordstyle=[3]\color{Orange},
	keywords=[4]{grb_rbgs_forward},
	keywordstyle=[4]\color{OliveGreen},
	breaklines=true,
	mathescape=true,
	tabsize=2,
	showstringspaces=false,
	numbers=left,
	xleftmargin=2em,
	stepnumber=1,
	numberstyle=\ttfamily\footnotesize,
	float=t,
	sensitive=true, % keywords are not case-sensitive
	morecomment=[l]{//}, % l is for line comment
	morecomment=[s]{/*}{*/}, % s is for start and end delimiter
	morestring=[b]" % defines that strings are enclosed in double quotes
}

\usepackage[noabbrev,capitalize]{cleveref}
\newacronym{hpc}{HPC}{High-Performance Computing}
\newacronym{fem}{FEM}{Finite Elements Method}
\newacronym{ai}{AI}{Artificial Intelligence}

\newacronym{hpl}{HPL}{High Performance Linpack}
\newacronym{pcg}{PCG}{Preconditioned Conjugate Gradient}
\newacronym{hpcg}{HPCG}{High Performance Conjugate Gradient}

\newacronym{pde}{PDE}{Partial Differential Equations}
\newacronym{cg}{CG}{Conjugate Gradient}
\newacronym{mg}{MG}{Multigrid}
\newacronym{gs}{GS}{Gauss-Seidel}
\newacronym{sgs}{SGS}{Symmetric Gauss-Seidel}
\newacronym{rbgs}{RBGS}{Red-Black Gauss-Seidel}

\newacronym{crs}{CRS}{Compressed Sparse Row}
\newacronym{lpf}{LPF}{Lightweight Parallel Foundations}
\newacronym{numa}{NUMA}{Non Uniform Memory Access}
\newacronym{amr}{AMR}{Algebraic Multigrid}
\newacronym{bsp}{BSP}{Bulk-Synchronous Parallel}

\newcommand{\alp}{ALP/GraphBLAS\xspace}

\newcommand{\grb}{GraphBLAS\xspace}

\newcommand{\code}[1]{\texttt{#1}}

\begin{document}

\title{
Effective implementation of the High Performance Conjugate Gradient benchmark on \grb
}

\author{
	\IEEEauthorblockN{1\textsuperscript{st} Alberto Scolari}
	\IEEEauthorblockA{
		\textit{Computing Systems Laboratory}\\
		{Zurich Research Center}\\
		{Huawei Technologies, Switzerland}\\
		%\textit{Huawei Zurich Research Center}\\
		%Zurich, Switzerland\\
		alberto.scolari@huawei.com
	}
	\and
	\IEEEauthorblockN{2\textsuperscript{nd} Albert-Jan Yzelman}
	\IEEEauthorblockA{
		\textit{Computing Systems Laboratory}\\
		{Zurich Research Center}\\
		{Huawei Technologies, Switzerland}\\
		albertjan.yzelman@huawei.com
	}
	%\and
}

\maketitle

\begin{abstract}
%This document is a model and instructions for \LaTeX.
%This and the IEEEtran.cls file define the components of your paper [title, text, heads, etc.].
%*CRITICAL: Do Not Use Symbols, Special Characters, Footnotes, or Math in Paper Title or Abstract.
%\lipsum[1]
Applications in \gls{hpc} environments face challenges due to increasing complexity.
Among them, the increasing usage of sparse data pushes the limits of data structures and programming models and hampers the efficient usage of existing, highly parallel hardware.
The \grb specification tackles these challenges by proposing a set of data containers and primitives, coupled with a semantics based on abstract algebraic concepts: this allows multiple applications on sparse data to be described with a small set of primitives and benefit from the many optimizations of a compile-time-known algebraic specification.
Among \gls{hpc} applications, the \gls{hpcg} benchmark is an important representative of a large body of sparse workloads, and its structure poses several programmability and performance challenges.
This work tackles them by proposing and evaluating an implementation on \grb of \gls{hpcg}, highlighting the main changes to its kernels.
The results for shared memory systems outperforms the reference, while results in distributed systems highlight fundamental limitations of \grb-compliant implementations, which suggests several future directions.
\end{abstract}

\begin{IEEEkeywords}
hpcg,multigrid,red-black gauss-seidel,graphblas
\end{IEEEkeywords}

\glsresetall

\section{Introduction}
The \gls{hpc} landscape is currently facing challenges due to the steadily increasing complexity of applications and platforms.
Applications from both classical scientific domains like physics and newer domains like data analytics or \gls{ai} require attaining high performance metrics while at the same time keeping programming efforts under control.
An important trend is the adoption of \emph{sparse} data, which model some real-world scenarios where certain entities are associated to others and only a small subsets of all possible associations is of interest.
Examples of this trend are diverse, ranging from \glspl{fem}-based applications for physics \cite{fem_intro} to graph-based applications for the analysis of large data sets \cite{graph_analysis}.
Computing platforms are also becoming increasingly complex, and writing reliable and performing applications requires nowadays costly expertise and platform-specific tools.
Current software solutions also fall short in adapting to the growing landscape of \gls{hpc} applications while ensuring effective hardware usage.
%, and the research on this topic is very lively.
%A well-known example of this trend are the BLAS routines \cite{blas_intro}, which the main hardware vendors must update on each new family product and which were originally designed for non-sparse (i.e., \emph{dense}) algebraic applications, much like the hardware platforms they run on.

In this scenario, applications of special interest in \gls{hpc} are iterative solvers for sparse linear systems \cite{sparse_solvers}, which allow studying many physical problems by relying on a sparse description of the physical system.
Several software packages implement sparse solvers on different hardware platforms \cite{sparse_solvers_sw}, and the research is increasingly active.
As a consequence, the need for standardization and comparison led to the introduction of a widely-accepted \gls{hpc} benchmark named \gls{hpcg} \cite{hpcg_intro}, from which the eponymous \emph{HPCG} rank for supercomputers was created \cite{hpcg_rank}.
\Gls{hpcg} solves a sparse linear problem $ A x = b $, where $ A $ is a sparse matrix of size $ n \times n $, $ b $ a dense, known vector and $ x $ the unknown vector.
Unlike dense solvers such as the well-known \gls{hpl} suite \cite{hpl}, \gls{hpcg} fully exploits the sparsity of the $ A $ matrix by implementing an iterative solver algorithm whose time and space complexity are proportional to the number of non-zeroes in $ A $ and by using suitable data structures.
Although \gls{hpcg} solves a specific physical problem, the algorithms it is composed of represent a wide class of solvers, and the ranking guidelines disallow problem-specific optimizations \cite{hpcg_tech_spec}.

Despite standardization efforts, multiple data structures for sparse data exist, whose choice depends on various factors (sparsity pattern, computation, hardware, etc.); as a consequence, the implementation of algorithms may strongly vary, leading to a complex design space.
Instead, productivity calls for an abstract representation of the computation as close as possible to its algebraic formulation: this led to the introduction of \grb \cite{grb}, a specification to develop algorithms for graphs with strong algebraic foundations.
Unlike BLAS  \cite{blas_intro} and other frameworks, \grb does not prescribe a specific storage format for vectors and matrices, leaving the choice to the specific implementation.
This abstraction enables domain experts to develop algorithms that are portable and leave optimizations to the specific implementation, thus achieving a complete separation of concerns.

Despite the increasing adoption of \grb, sparse solvers are still an area of active development and challenge both the \grb specification and its implementations.
This paper presents an implementation of \gls{hpcg} on top of \grb, addressing the related design and implementation choices.
Hence, it brings the following contributions:
\begin{itemize}
\item it discusses the main design decisions to implement \gls{hpcg} on top of \grb, showing how its kernels can be implemented on top of opaque data structures
\item it assesses the performance of the implementation in shared memory systems typical of \gls{hpc} environments, outperforming the reference with no manual or architecture-specific optimization efforts
\item it assesses the performance of the implementation in a distributed system, showing current limitations
\item it discusses theoretical limitations and possible advancements to the current implementations and to the \grb specification
\end{itemize}

This paper is organized as follows.
\Cref{sec:background} introduces to the background information, in particular the problem \gls{hpcg} solves, its main kernels and the \grb specification.
\Cref{sec:design} discusses the design decisions necessary to implement \gls{hpcg} on top of \grb.
\Cref{sec:impl} explains the main characteristics of the implementation, while \cref{sec:experiments} shows its evaluation on several \gls{hpc}-grade systems.
\Cref{sec:related} highlights the main efforts to accelerate \gls{hpcg} on \gls{hpc} architectures.
Finally, \cref{sec:concl} discusses the achievements of this work and possible future work.

\section{Background} \label{sec:background}
This section explains the structure of the \gls{hpcg} benchmark, its kernels and performance characteristics.
It then describes the aspects of \grb that are relevant to this paper.

\subsection{The HPCG benchmark} \label{sec:benchmark_kernels}
Many problems of interest to physicists and engineers are represented by \glspl{pde} and model the problem space as a 1D, 2D or 3D grid.
Elliptic \glspl{pde} are an important class, as they describe many common problems and are usually discretized on a 3D spatial grid of $ n_x \times n_y \times n_z = n $ points, where the interactions are modeled according to the physics of the system.
These interactions often involve only points that are spatially close (e.g., within a halo), producing inherently sparse matrices.

\Gls{hpcg} \cite{hpcg_tech_spec,hpcg_github} solves a heat-diffusion problem on a semi-regular 3D grid via a preconditioned \gls{cg} solver \cite{cg_intro}, modeling heat exchanges among neighboring elements.
\gls{hpcg} main kernels are:
\begin{enumerate}
\item input generation
\item \glsentrylong{cg} solver, in turn relying on the
\item \glsentrylong{mg} preconditioner, implemented on top of the
\item \glsentrylong{sgs} smoother;
\item restriction and refinement, to recursively call the \gls{mg} preconditioner and use its result;
\end{enumerate}
The following sections delve into the details of these kernels.

\subsection{Input generation}
An iterative sparse solver has three inputs: the system matrix $ A $, the right-hand side vector $ b $ and the initial hint for the solution $ x^{(0)} $.
Since \gls{hpcg} simulates a well-known heat diffusion problem on a semi-regular grid, these inputs are generated automatically, with $ b = 1 $ and $ x^{(0)} = 0 $.

\subsection{Conjugate Gradient solver}
The \gls{cg} solver \cite{cg_intro} is a generic, iterative algorithm for the solution of \glspl{pde}, which starts from the user-given solution $ x^{(0)} $ and iteratively refines it until converging to the final solution $ x^{(i)} $ after $i$ iterations, or until the solution is ``close enough'', i.e., until the module of the \emph{residual vector} $ r = b - A x^{(i)} $ is smaller than a given threshold.
The main computational kernels of \gls{cg} are the sparse vector product \code{spmv}, the dot product of two vectors \code{dot} and the linear combination of two vectors \code{waxpby}.
Among them, \code{spmv} dominates the runtime due to its complexity: \gls{hpcg} system matrix $A$, which represents a semi-regular 3D grid with halo 1, has from 8 to 27 nozeroes per rows, resulting in a total size $\Theta \left( n \right) $.
Therefore, the work complexity of an \code{spmv} operation in \gls{hpcg}, and thus of an iteration of the \gls{cg} solver, is $\Theta \left( n \right) $.

\subsection{\acrfull{mg} preconditioner}
\begin{lstlisting}[language = pseudo,caption={Steps of the \acrshort{mg} preconditioner},label={lst:mg}]
function MG(mg_level,$z$,$r$)
	$z \leftarrow$ sgs_smoother(mg_level,$z$,$r$)
	if (mg_level.coarser_sys = nil)
		return $z$
	$f \leftarrow$ mg_level.$A$ * $z$
	$ r_c $ $\leftarrow$ restrict(mg_level,$r - f$)
	$ z_c \leftarrow 0 $
	$ z_c \leftarrow $ MG(mg_level.coarser_level,$z_c$,$r_c$)
	$ z \leftarrow z $ + refine(mg_level,$z_c$)
	$z \leftarrow$ sgs_smoother(mg_level,$z$,$r$)
	return $z$
\end{lstlisting}
The second main kernel of \gls{hpcg} is the \gls{mg} preconditioner \cite{mg}, which improves the convergence towards the solution at the cost of higher complexity and slightly less generalization w.r.t. the \gls{cg} solver.
This preconditioner improves the initial solution by computing an approximation $ z $ of $ A^{-1} r $ in a recursive fashion, which then contributes to the main solution of \gls{cg} $ x^{(i)} $.
It does so by attempting to solve a problem $ A z = r $ through successive, recursive restriction and refinement steps: on each recursion the current solution $ z $ (initially $ 0 $) is improved by a smoothing step and by solving a smaller problem based on a restriction of $ r $ and $ z $; the result of this recursive problem is then refined and contributed back to $ z $.

\Cref{lst:mg} sketches the \gls{mg} preconditioner in more detail:
\begin{enumerate}
\item \label{item:presmoothing} \textbf{line 2}: smooth $ z $ from $ r $ via the \gls{sgs} smoother, so that $ A z \approx r $;
\item \label{item:base_case} \textbf{lines 3-4}: if the maximum number of coarsening levels has been reached, simply return the current solution $ z $;
\item \label{item:residual} \textbf{line 5}: compute $ f \leftarrow A z $, i.e., the current residual for the problem $ A z = r $;
\item \textbf{line 5}: compute the restricted residual $ r_c $ for the problem $ A z = r $, i.e., the restriction of $ r - f $;
\item \label{item:recursive} \textbf{line 8}: improve the initial $ z_c \leftarrow 0 $ (line 6) by recursively calling the \gls{mg} preconditioner on $ r_c $ and the coarser system \emph{mg\_level.coarser\_level};
\item \textbf{line 9}: refine $ z_c $ to the current system and add it to $ z $, thus improving on the initial solution $ z $;
\item \textbf{line 10}: smooth $ z $ again to remove oscillatory components added in the previous step.
\end{enumerate}

\subsection{\acrfull{sgs} smoother} \label{par:sgs_smoother}
The basic \gls{gs} algorithm smooths $ z $ by updating each value $z_i$ of it as
\begin{equation} \label{eq:gs_basic}
z_i \leftarrow \left( A_{i,i} \right)^{-1} \left( r_i - \sum_{j = 0, j \ne i}^{n-1} A_{i,j} z_j \right) \quad 0 \leq i < n;
\end{equation}
essentially, \cref{eq:gs_basic} solves the $ i $-th equation of the problem $ A z = r $.
\Gls{gs} prescribes the order of increasing $ i $ to compute \cref{eq:gs_basic}, causing the output value $ z_i $ to depend on the input values $ z_j$, $ j < i $ computed in the previous iterations of \cref{eq:gs_basic} and for which $ A_{i,j} \ne 0 $; hence, $ A_{i,j} \ne 0 $ determines the presence of direct $ (i,j) $ dependencies, which then propagate transitively across values depending on $A$.
Therefore, the available parallelism depends on $ A $ itself.
For \gls{hpcg}, $A$ is a semi-regular grid over a homogeneous spatial domain, which causes $ (i,j) $ dependencies to propagate transitively from \emph{all} $ j < i $, making the execution of \gls{gs} in \gls{hpcg} inherently sequential, a major performance bottleneck in \gls{hpc} highly parallel systems.

The \gls{sgs} version adds to the previous step (called \emph{forward sweep}) another step, called \emph{backward sweep}, which performs the same operations but with $ i $ from $ n - 1 $ to $ 0 $, improving the convergence for many problems \cite{mg}; nonetheless, the aforementioned performance issue holds for \gls{sgs} as well.

The complexity of \gls{sgs} in \gls{hpcg} also depends on the structure of $A$, because \cref{eq:gs_basic} is computed $n$ times for each row $i$ of $A$.
Since each row has $ \Theta \left( 1 \right) $ nozeroes, the total work complexity of \gls{sgs} is $\Theta \left( n \right) $.

\subsection{Restriction and refinement} \label{sec:restriction_refinement}
The restriction operation projects a vector from a fine space to a coarse space, the refinement operation does the opposite.
Restriction occurs to create the inputs for the coarser problem (line 6 of \cref{lst:mg}), whose output is refined to improve the current solution (line 9).

\Gls{hpcg} restriction operation implements \emph{straight injection} \cite{mg_injection}: it restricts the vector in the fine space by a factor of two along each dimension and populates each point of the coarse vector with the value at the lowest coordinates of the corresponding octet in the fine space.
Correspondingly, the refinement operation populates the finer vector with the corresponding values of the coarse vector and zeroes elsewhere.

As these operations move data in a fixed sparse pattern and do not require computation,
the \gls{hpcg} reference implementation performs it \emph{in-place} by directly accessing the input and output arrays, thus making assumptions on the data structure.

\Gls{hpcg} 3D input domain requires restricting/refining vectors by a factor eight; furthermore, the number of restriction/refinement levels is limited to a fixed number (typically 4), for a total work complexity of $\Theta \left( n \right) $.
Hence, the whole work complexity of \gls{hpcg} reference implementation is $\Theta \left( n \right) $.

\subsection{Data distribution for distributed execution} \label{sec:ref_distrib}
A focal design aspect of the reference \gls{hpcg} implementation is the data distribution across nodes for distributed execution.
To achieve minimal data exchange, the reference optimally splits the physical simulation grid of sizes $ n_x \times n_y \times n_z = n $ across the $ p $ nodes.
To achieve optimality, it computes the factorization $  p = p_x \times p_y \times p_z $ that maps optimally to the $ n_x \times n_y \times n_z $ grid.
In this way only the 2D halos in the physical space need to be exchanged among nodes: if each local dimension is $ s_d = n_d / p_d $, $ d \in \{x,y,z\} $, the number of points each node exchanges with its eight 3D neighbors is $ h = 2 \left( s_x s_y + s_y s_z + s_x s_z \right) $.
The rows of matrix $ A $ are partitioned following this distribution and nodes update only vector data during the computation.
Hence, if $ n_d = \Theta \left( \sqrt[3]{n} \right) $ and $ p_d = \Theta \left( \sqrt[3]{p} \right) $ with $ d \in \{x,y,z\} $, then each nodes exchanges $ \Theta\left( \sqrt[3]{n^2 / p^2} \right) $ values before an \code{mxv} operation.

\subsection{The \grb specification and its implementations}
\grb \cite{grb} is a specification based on two core concepts: graphs can be represented via matrices and algorithms on graphs can be expressed via algebraic operations defined over algebraic structures like monoids and semirings.
These abstractions ease programming while providing with solid assumptions to carry optimizations, like the properties of the algebraic structure.
This specification stresses the importance of sparse data structures and operations, and is thus a good fit for modern \gls{hpc} applications and for sparse linear solvers in particular, inasmuch it allows domain experts to implement algorithms without extensive programming and optimization efforts, while ensuring good performance thanks to its performance semantics.
From a high-level perspective, a \grb program is thus a sequence of standard algebraic operations like \code{spmv}, \code{dot} and \code{waxpby} performed on opaque data structures representing matrices and vectors, augmented with information about the algebraic structure.

Containers opaqueness allows implementations the freedom to make different choices based on the platform and the data, which determines performance.
Multiple implementations of the official C specification \cite{grb_spec} exist, like SuitSparse:\grb \cite{suitesparse}, vendor-specific ones \cite{ibmgraphblas} and also C++ implementations \cite{gbtl,alp_paper}, which employ template meta-programming to optimize operations based on the user-given algebraic structure.

\section{Design of HPCG kernels for \grb} \label{sec:design}
This section explains the design decisions to implement \gls{hpcg} on \grb.
Some \gls{hpcg} kernels in \cref{sec:benchmark_kernels} require little effort, because the operations in \grb closely match those in the reference \gls{hpcg} implementation.
This is the case for the \gls{cg} solver, which is composed of standard algebraic operations that easily map to \grb, keeping the same work complexity $\Theta \left( n \right) $ of the reference.
%The input generation also closely follows the original specification, building the system matrix $ A $ from the triplets of non-zeroes generated as in the reference implementation.
%
Instead, the components of the \gls{mg} preconditioner do not straightforwardly map.
The following sections explain their design.

\subsection{\acrfull{sgs} smoother}  \label{sec:smoother_design}
\begin{lstlisting}[language = pseudo,caption={\grb pseudo-code for the \acrshort{rbgs} forward pass},label={lst:grb_rbgs_smooth}]
function grb_rbgs_forward(mg_level,$z$,$r$,$c$)
	for $k$ from $0$ to $c$
		$ s \leftarrow $ mxv(mg_level.$colors$[$k$],
			mg_level.$A$, $z$)
		apply(i -> {$z[i] \leftarrow $ ($r$[$i$] - $s$[$i$] +
				$z$[$i$] * mg_level.$A\_diag$[$i$]) /
				mg_level.$A\_diag$[$i$]},
				mg_level.$colors$[$k$])
	return $z$
\end{lstlisting}
As from \cref{par:sgs_smoother}, \gls{hpcg} smoother is, in combination with its input, inherently sequential.
Furthermore, the ordering of updates to $ z_i $ in \cref{eq:gs_basic} makes this kernel not easily expressible with standard algebraic operations.
One could attempt to workaround this latter limitation, e.g., by implementing the \gls{sgs} kernel with $ n $ \code{spmv} operations on $ n $ vectors and updating one point at a time, but the time and memory overheads of such a solution would be prohibitive, and the execution would still be sequential.

Instead, multiple works \cite{arm_hpcg,k_supercomp_hpcg} replace the \gls{sgs} smoother with \gls{rbgs}, which overcomes these limitations by relaxing some $ (i,j) $ dependencies via a partitioning scheme that groups together indices.
Indeed, \gls{hpcg} technical specification \cite{hpcg_tech_spec} allows changing the smoother under the condition that the new one passes the internal symmetry test.
\Gls{rbgs} relaxation enables updating independent indices in parallel at the cost of a higher number of iterations to achieve the same smoothing effect of \gls{gs} \cite{rbgs_effect}.
Nonetheless, the performance gain thanks to parallelisation usually outweighs the higher number of iterations.

If the partitioning scheme finds $ c $ colors $ C_k $, $ 0 \leq k < c $, all $ z_i $ for which $ i \in C_k $ can be updated in parallel as in \cref{eq:gs_basic}.
This allows implementing \gls{rbgs} with natively parallel \grb primitives, as \cref{lst:grb_rbgs_smooth} sketches: here, operations on all elements of a color $k$ are implemented with \grb primitives, thus inherently parallel; instead, colors are processed sequentially (loop of line 1) to meet the dependencies among them.
The \emph{masked} \code{mxv} of line 3 implements the summation of \cref{eq:gs_basic} for all values of color $k$: the first argument \code{mg\_level.$colors$[$k$]} is a binary mask marking the indices $ i \in C_k $, so that \code{mxv} computes the outputs only for these positions into \code{s}.
The following \code{apply} operation applies the lambda in the first argument in parallel for all marked elements of the mask, i.e., for all $ i \in C_k $, computing the remaining part of \cref{eq:gs_basic} and then updating $ z_i $ as a side effect.
This function needs to access the diagonal value of $ A $, which is stored in a dedicated vector \code{mg\_level.$A\_diag$} since \grb does not allow accessing individual matrix values in constant time.
This vector is initialized during input generation.

%I moved this paragraph forward as I felt like the text flows better as a result ALBERTO_PLEASECHECK
An effective partitioning scheme depends on $ A $: the larger the parts of a partitions, the larger the exposed parallelism.
For \gls{hpcg}, an effective and simple partitioning scheme can be implemented via a \emph{greedy} coloring algorithm \cite{coloring} that tracks only direct $ (i,j) $ dependencies: if there is a \emph{direct} $ (i,j) $ dependence, than $ i $ and $ j $ must have different colors.
Higher parallel efficiency is achieved when the number of colors is low, i.e., the coloring is near-optimal.
For \gls{hpcg} 3D grid, greedy coloring achieves optimal results, finding eight colors.

Partitioning schemes other than our greedy one are possible.
Some of them also group the indices to improve the spatial locality of data accesses \cite{arm_hpcg,k_supercomp_hpcg}, so that indices having the same color $ C_k $ are stored in physically contiguous positions.
The choice of a more advanced partitioning scheme depends on the physical domain and is thus orthogonal to the programming framework, and hence outside the scope of this work.
In the reference implementation, any grouping of indices can be realized by manipulating the data structures of the system $ A $, which are non-opaque, while \grb makes this impossible because of containers opaqueness.
However, a grouping scheme can be performed via a row \emph{permutation} matrix $ P $ and its transposed $ P^{T} $ as $ P^{T}AP $, which is supported in \grb via the \code{mxm} operation.

Since \gls{rbgs} is a relaxation of \gls{gs} to expose more parallelism, its work complexity also takes into account the number of processing units $p$.
With serial execution the work complexity is still $\Theta \left( n \right) $, because it performs the same computation of \cref{eq:gs_basic}, though in a different order.
Under the assumption that all partitions are large enough to efficiently parallelise their computation using $p$ cores, the expected speedup is $p$.
Since $n$ is orders of magnitudes higher than $p$ in realistic workloads and the coloring algorithm we use here can optimally identify the eight colors for \gls{hpcg} input domain, we indeed assume this assumption holds and that \gls{rbgs} has work complexity $ \Theta \left( n / p \right) $.

\subsection{Restriction and refinement} \label{sec:restriction_design}
The \Gls{hpcg} reference implementation directly accesses the storage to compute both restriction and refinement, which is impossible in \grb because vectors are opaque objects.
%Therefore, these operations must be implemented via algebraic primitives.
As \gls{hpcg} \emph{straight injection} restriction is a linear operation, it corresponds to a matrix--vector multiplication with a rectangular matrix of sizes $ n \times \frac{n}{8} $ and naturally is expressible in \grb.
Similarly, \gls{hpcg} refinement is the transposed operation of restriction.
Therefore, the work complexity is that of an \code{mxv} operation on \gls{hpcg}, i.e., $ \Theta \left( n \right) $.

This constraint forces the materialization of a matrix, whereas \gls{hpcg} reference implements restriction as an array of indices to copy from the fine vector to the coarse one, and uses the same vector to implement refinement.
Because of the data structures typically used to describe sparse matrices, e.g., the three arrays of \gls{crs} \cite{crs}, a \grb implementation of refinement/restriction may incur additional storage and time overheads compared to the reference; the work complexity, however, is the same.

%I disabled this paragraph ALBERTO_PLEASECHECK
%Another limitation of this approach is that only linear operations can be expressed, which may exclude some highly-nonlinear problems requiring nonlinear restriction/refinement.
%However, this limitation is out of the scope of this work, which deals only with linear systems.

%
\begin{lstlisting}[language = {my_cpp},caption={\alp C++ implementation of the forward \gls{rbgs} pass. Error checking omitted.},label={lst:grb_rbgs_smooth_impl}]
void grb_rbgs_forward(
	const Matrix<double> &A,
	const Vector<double> &A_diag,
	const std::vector<Vector<bool>> &colors,
	const Vector<double> &r,
	Vector<double> &x,
	Vector<double> &tmp, // workspace buffer
	Ring &ring
) {
	for(size_t c = 0;c < colors.size();++c){
		mxv<descriptors::structural>(
			tmp, colors[c], A, x, ring);
		eWiseLambda( [&](size_t i) {
				double d = A_diag[i];
				double v = r[i]-tmp[i]+x[i]*d;
				x[i] = v/d; },
			colors[c], x, r, tmp, A_diag);
	}
}
\end{lstlisting}

\newcommand{\alpi}{ALP\xspace}
\newcommand{\refi}{Ref\xspace}

\section{Implementation} \label{sec:impl}
% This section shows the most important implementation details of the kernels discussed in \cref{sec:design}.
We implement \gls{hpcg} on \grb using a C++ variant of the \grb specification by Yzelman et al.\ named \alp \cite{alp_paper}, which describes \grb primitives and their algebraic structure via C++ templates to leverage the (algebraic) type safety checks and the automatic performance optimization of modern compilers via template meta programming-- while keeping a concise code base.
From here on, we call this implementation \emph{\alpi}.
%Thanks to this choice, one can expect
We expect \alpi to attain performance comparable to that of a hand-coded C/C++ application such as the \gls{hpcg} reference code, which is also coded in C++ that, similar to \alp, does not employ abstractions that may prevent compiler optimizations, %such as virtual methods, raw pointers, etc., -- NOTE disabled due to originally being in parenthesis and seems not strictly necessary ALBERTO_PLEASECHECK
thus enabling a fair comparison of the two implementations.
\Cref{lst:grb_rbgs_smooth_impl} implements the forward pass of \gls{rbgs} with \alp, showing in red and orange the \grb containers and primitives, respectively.
It closely follows the structure of \cref{lst:grb_rbgs_smooth},
%ALBERTO_PLEASECHECK -- the below is unclear to me and probably needs rephrasing
showing the \code{A},\code{A\_diag},\code{colors} fields of \code{mg\_level} in \cref{lst:grb_rbgs_smooth} as explicit C++ function parameters:
here, the color masks are passed via a common C++ data structure storing the boolean masks of type \code{Vector<bool>}.

\alp provides several \emph{descriptors} to pass domain information to operations.
In line 11 of \cref{lst:grb_rbgs_smooth_impl} \alp \emph{structural} descriptor leverages the optimized implementation of \code{Vector} for sparsity, causing \code{mxv} to ignore the actual nonzero value of the color mask \code{colors[c]} and follow only the sparsity pattern, avoiding useless memory accesses.
Another example is the refinement operation of \cref{sec:restriction_design}, which uses the restriction matrix transposed; to avoid materializing the transposed matrix, the \code{mxv} operation for refinement is added the \emph{transpose\_matrix} descriptor to use the restriction matrix directly.
% \usepackage{booktabs}

\renewcommand{\arraystretch}{1.4}

\begin{table}
\caption{\acrshort{bsp} asymptotic cost components for the distributed implementations.}
\centering

\begin{tabular}{@{}ccc@{}}
\toprule
				& \refi                     & \alpi                         \\ \midrule
computation     & $ n / p $            & $ n / p $ \\
communication   & $ \sqrt[3]{n^2 / p^2} $ & $ n / p (p - 1) \approx n $ \\
synchronization & $ 1 $       & $1$                      \\ \bottomrule
\end{tabular}

\label{tab:bsp_costs}
\end{table}

\renewcommand{\arraystretch}{1}

\alp also provides several \emph{backends}, i.e. implementations of containers and operations tailored to different execution environments that can be chosen at compile time and that all meet \grb performance semantics.
Throughout our experiments, \alpi uses the \emph{shared-memory} backend for multicore machines and the \emph{hybrid} backend for distributed and multicore execution.
The shared-memory backend distributes the execution among cores using OpenMP threads, taking into account \gls{numa} architectures via a \gls{numa}-aware memory allocator \cite{libnuma} and preferring interleaved allocations to balance memory accesses among domains and avoid bottlenecks.
The hybrid backend relies on the \gls{lpf} communication layer \cite{lpf} to provide high-performance communication functionalities.
The \gls{lpf} interface and semantics abide by the \gls{bsp} \cite{bsp_model} model of computation,
while the hybrid \alp backend assumes a 1D grid of nodes and splits matrix rows and vectors according a block-cyclic distribution.
This requires all nodes to have an up-to-date copy of the entire input vector before an \code{mxv} operation, which is enforced via a sparsity- and mask-aware all-to-all communication.
The communication cost for an \code{mxv} with dense vectors is proportional to the maximum amount of data any node sends or receives; i.e., $ \Theta \left( n (p - 1) / p \right) \approx \Theta \left( n \right) $.
Since there is a constant number of \code{mxv} operations within \gls{hpcg} and the number of benchmark iterations is controlled by the user, the total communication complexity of \alpi is $ \Theta \left( n \right) $.
Similarly, synchronizations occur before every \code{mxv} operation, thus with a global synchronization cost of $ \Theta \left( 1 \right) $.

The linear complexity of communication costs are expected to affect performance, as experiments in \cref{sec:experiments} confirm.
\Cref{sec:concl} further elaborates on this limitation, highlighting inherent limits of \grb and discussing mitigations.

We implement the \gls{rbgs} smoother of \cref{sec:smoother_design} also in the reference \gls{hpcg} code base available online \cite{hpcg_github}.
From here on, we call this implementation \emph{\refi}.
% ALBERTO_PLEASECHECK: does "It" below refer to the RBGS smoother, the HPCG code base, \refi only, or all of these?
\refi uses OpenMP to update the values within a single color in parallel in a shared memory system, together with MPI-3 primitives to exchange information among nodes for distributed execution; in particular, MPI \code{Irecv}/\code{Isend} are used to overlap execution and communication when updating the values of $x$ for each color.
% ALBERTO_PLEASECHECK: new sentence below to clarify what we discussed last week
Unlike \refi, the hybrid \alp backend uses blocking \grb semantics and thus features no such overlap.
Like \alpi (\cref{sec:smoother_design}), however, also \refi has $ \Theta \left( n / p \right) $ work complexity.

% ALBERTO_PLEASECHECK: I added "in \refi" below, that's correct right?
The color-aware data exchange in \refi follows the same design principles in \cref{sec:ref_distrib}, where each node exchanges with the geometrical neighbors only the values in its halo having the specific color to be updated.
This keeps the same communication complexity derived in \cref{sec:ref_distrib}: $ \Theta \left( \sqrt[3]{n^2 / p^2} \right) $.
In \refi, neighboring nodes synchronize after updating each color.
Since the number of colors found for \gls{hpcg} does not depend on $n$ or $p$, the total synchronization is again $ \Theta \left( 1 \right) $.
\Cref{tab:bsp_costs} summarizes the \gls{bsp} asymptotic cost components of both \alpi and \refi.

As in the original code base, \refi does not use any \gls{numa}-aware memory allocator, thus resorting to the standard domain-local memory allocations of the C++ standard library implementation.
% ALBERTO_PLEASECHECK: I added in the below, again to emphasize the contrast (similar to blocking vs. nonblocking/overlap)
The shared-memory and hybrid \alp backends, however, do perform \gls{numa}-aware allocations.

\section{Experimental Results} \label{sec:experiments}

\newcommand{\intels}{x86\xspace}
\newcommand{\arms}{ARM\xspace}

\newcommand{\maxnodes}{7\xspace}

% \usepackage{booktabs}
\begin{table}
\caption{Details of the experimental machines; \emph{italics} means per socket, otherwise per node.}

\begin{tabular}{@{}ccc@{}}
\toprule
						& x86             & ARM              \\ \midrule
\emph{CPU}                     & Xeon Gold 6238T & Kunpeng 920-4826 \\
\emph{cores}                   & 22              & 48               \\
\emph{threads}                 & 44 (HT enabled) & 48               \\
\emph{max frequency (GHz)}     & 3.70            & 2.6              \\
\emph{L3 cache (MB)}           & 30.25           & 48               \\
\emph{per core L2 cache (KB)}           & 1024            & 512              \\
\emph{memory channels}         & 6               & 8                \\
\emph{NUMA domains }           & 1               & 2                \\
sockets                 & 2               & 2                \\
RAM memory (GB)         & 192             & 512              \\
max DDR frequency (MHz) & 2933            & 2933             \\
attained bandwidth (GB/s)  &  192  &  246.3   \\
\begin{tabular}[c]{@{}c@{}}network adapter,\\ bandwidth\end{tabular} &  & \begin{tabular}[c]{@{}c@{}}Mellanox ConnectX-5,\\ 2x100Gb/s\end{tabular} \\ \bottomrule
\end{tabular}

\label{tab:setup}
\end{table}
This section discusses the experiments to validate the two implementations proposed in \cref{sec:impl}.
%: the one implemented on \alp, here called \emph{\alpi}, and the one based on the reference implementation, here called \emph{\refi}.
\Cref{sec:exp_shmem} discusses the results on two systems: the first one is a dual-socket machine using an x86-64 architecture, here called \emph{\intels}; the second one is a dual-socket ARM64 machine, here called \emph{\arms}.
\Cref{tab:setup} shows the relevant hardware characteristics of both systems.
Both employ Ubuntu Linux 20.04 and employ GCC 9 as compiler.
Neither \refi nor \alpi employ any architecture-specific library or instruction, thus remaining portable.
An InfiniBand network interconnects up to \maxnodes \arms machines, providing a cluster for scale-out experiments that \cref{sec:exp_dist} discusses; there, \refi uses the optimized OpenMPI distribution provided with the network adapter.
%
%In the next sections and in the plots, the implementation this work proposes is called \emph{\alpi}, while the implementation in the reference code base is called \emph{\refi}.
Although multiple runs of the \gls{rbgs} may improve convergence, \gls{hpcg} performs by default one run, as this work does across all experiments.

All experiments achieve numerically comparable results, which allows fixing the number of iterations across all of them, thus making execution times directly comparable.
Each experiment is repeated 10 times and the plots in this section show the average value.
Unbiased standard deviations across experiments are negligible and thus not shown.

\subsection{Results on shared memory} \label{sec:exp_shmem}
\begin{figure}
\centering
\includegraphics{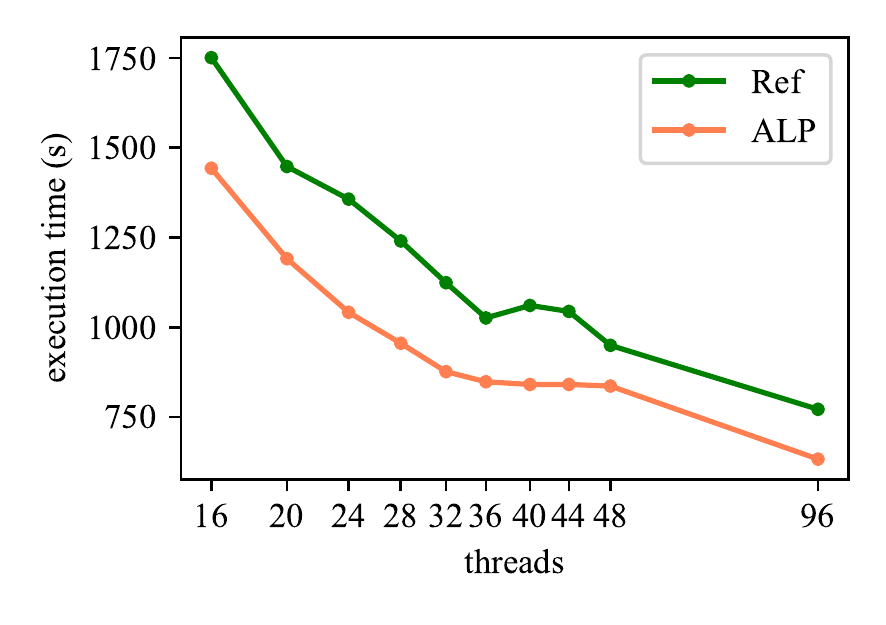}
\caption{Execution time of \alpi and \refi with increasing number of threads on \arms}
\label{fig:shmem_arm}
\end{figure}
\begin{figure}
\centering
\includegraphics{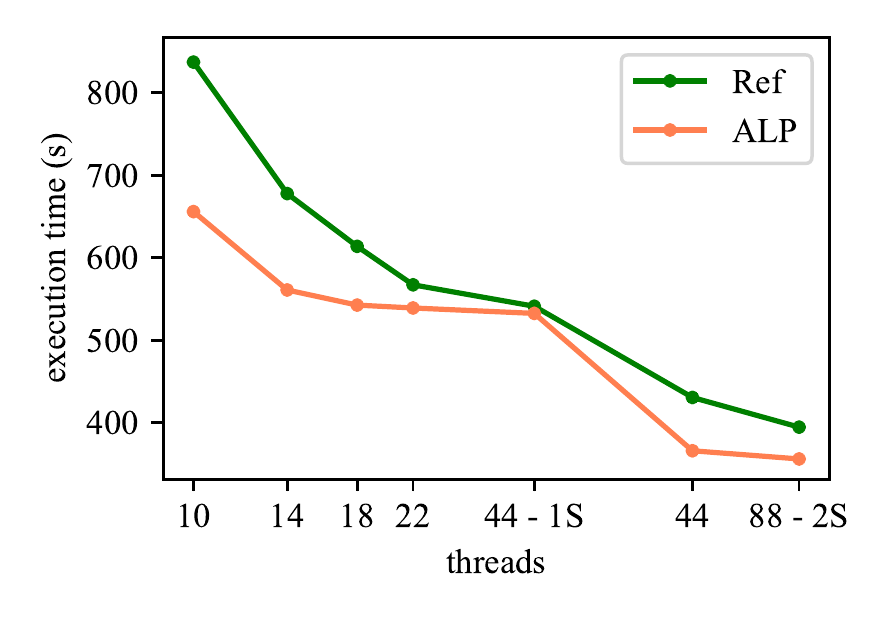}
\caption{Execution time of \alpi and \refi with increasing number of threads on \intels}
\label{fig:shmem_x86}
\end{figure}
\Cref{fig:shmem_arm} and \cref{fig:shmem_x86} show the strong scalability on both shared memory systems.
In both plots, the \emph{x} axis shows the number of application threads, increasing from less than those on a single socket to all those available on both sockets.
To avoid caching effects, the problem size is set to the maximum that fits in the system memory, which is higher for \arms: this motivates the different time ranges between the two plots.
Threads are pinned on physical cores, packing them on one socket if possible, i.e., with at most 48 threads on \arms and 22 threads on \intels.
\Cref{fig:shmem_x86} also shows results with 44 threads on one socket (value ``44 - 1S'' on the \emph{x} axis) and with 88 threads on two sockets (value ``88 - 2S'' on the x-axis), thus with hyperthreads.
For the \refi results on two sockets, we test in both systems the default allocation policy as well as the interleaved one,
enforced via the \code{numactl} command.
As the latter always outperforms the former, the plots report only the interleaved \refi execution--
%ALBERTO_PLEASECHECK is this reformulated OK? I want to prevent (like you did I think) the question why we're not just plotting both, i.e., because it's not interesting and should not matter :P
thus assuming that the exploration regarding \gls{numa} allocation is a common step for \gls{hpc} practitioners, especially when dealing with multi-socket configurations.

Both \cref{fig:shmem_arm} and \cref{fig:shmem_x86} show that \alpi outperforms \refi with all threads configurations.
When the threads fit in one socket, \alpi shows on both systems to saturate more quickly than \refi; this is due to the optimization available in \alpi, where the \grb semantics and the C++ template-based implementation allow propagating rich information across software layers and to the compiler.
In \cref{fig:shmem_arm}, the performance of \refi slightly degrades when the number of threads approaches the number of cores on single socket, which we attribute to the \gls{numa}-unaware allocation policy and to the two \gls{numa} domains per socket.
In \cref{fig:shmem_x86} with 44 threads pinned on a single socket (value ``44 - 1S''), the two implementations are close, showing that \refi saturates only with hyperthreading; this suggests that \alpi is more optimized.

\newcommand{\refres}{refinement/restriction\xspace}

\subsection{Results on the distributed system} \label{sec:exp_dist}
\begin{figure}
	\centering
	\includegraphics{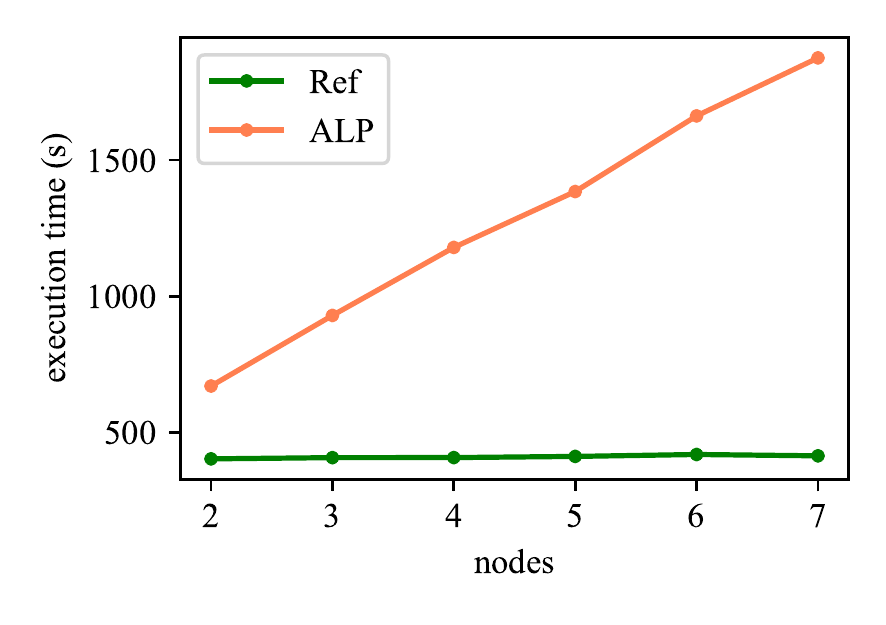}
	\caption{Execution time of \alpi and \refi with increasing number \arms nodes}
	\label{fig:dist}
\end{figure}
\Cref{fig:dist} shows the results from weak-scaling experiments on a test cluster of \arms systems.
These results confirm the asymptotic analysis in \cref{tab:bsp_costs}.
The \refi implementation shows a good weak scalability behavior, with execution times differing by at most 5\% among different number of nodes, while the input size $n$ grows proportionally to $p$.

Instead, the execution time of \alpi increases linearly with the number of nodes, conform to \cref{tab:bsp_costs}.

\subsection{Scalability of Refinement/Restriction and RBGS} \label{sec:mg_breakdown}
\begin{figure}
	\centering
	\includegraphics{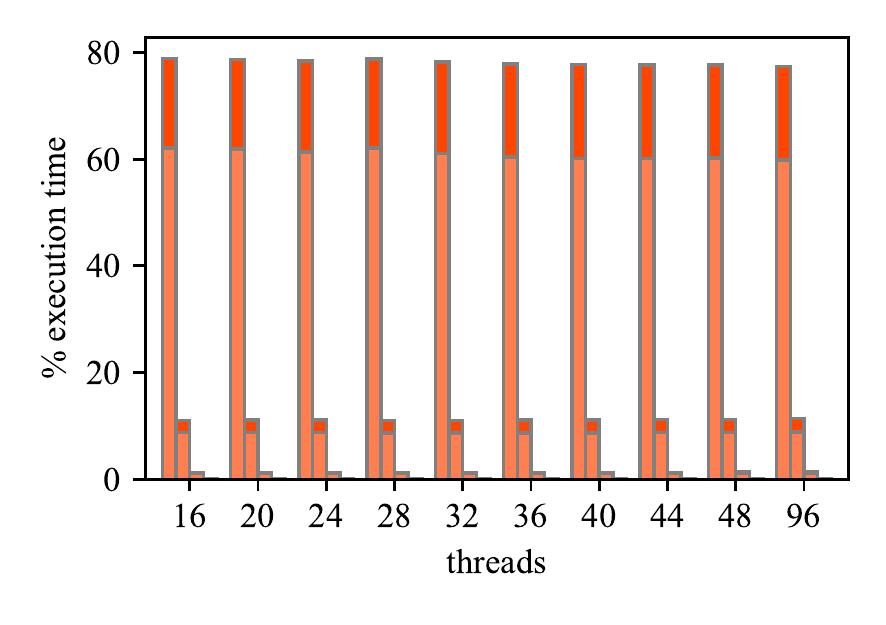}
	\caption{Percentage of execution time spent in \refres (dark) and in \acrshort{rbgs} (bright) for shared memory \alpi on \arms}
	\label{fig:shmem_arm_alp_mg}
\end{figure}
\begin{figure}
	\centering
	\includegraphics{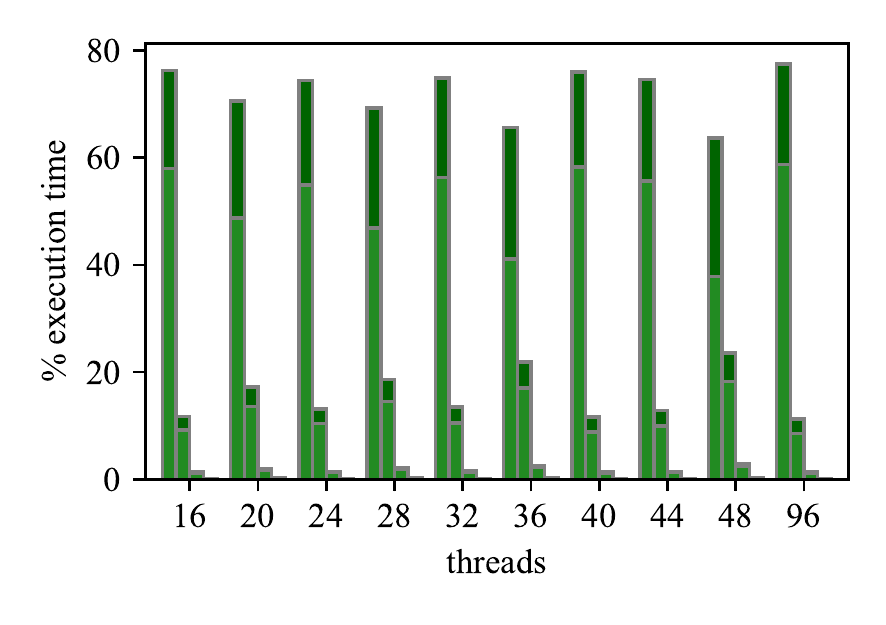}
	\caption{Percentage of execution time spent in \refres (dark) and in \acrshort{rbgs} (bright) for shared memory \refi on \arms}
	\label{fig:shmem_arm_ref_mg}
\end{figure}
\begin{figure}
	\centering
	\includegraphics{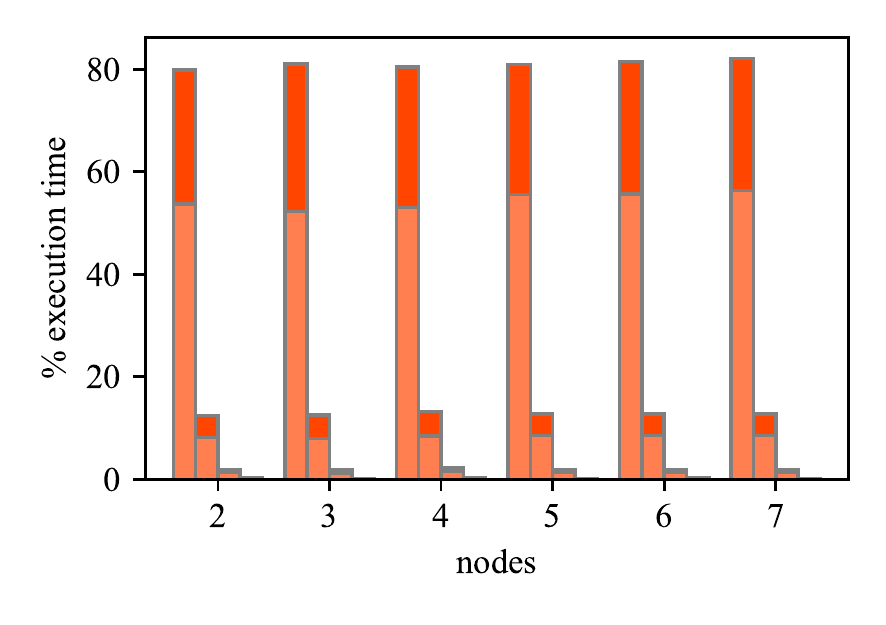}
	\caption{Percentage of execution time spent in \refres (dark) and in \acrshort{rbgs} (bright) for distributed \alpi}
	\label{fig:dist_alp_mg}
\end{figure}
\begin{figure}
	\centering
	\includegraphics{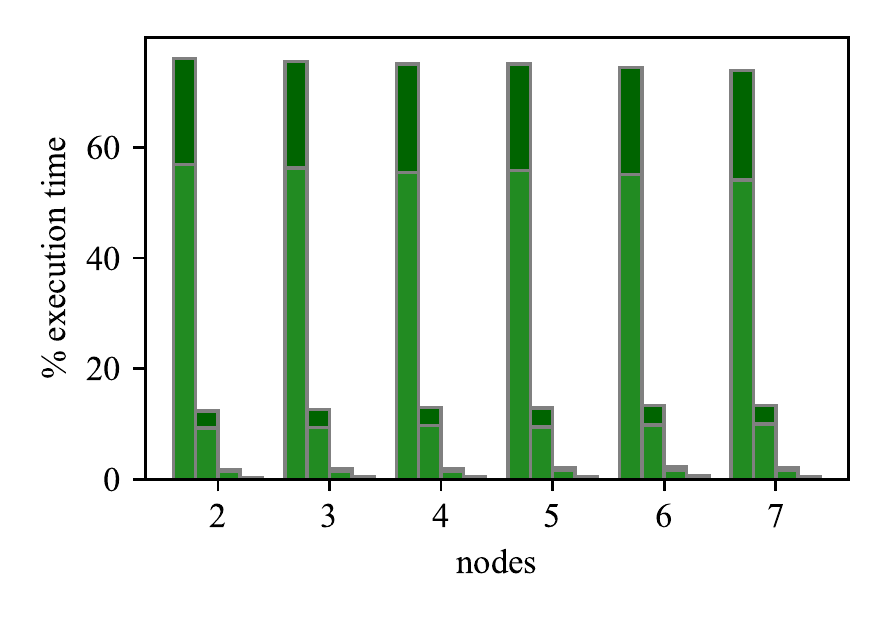}
	\caption{Percentage of execution time spent in \refres (dark) and in \acrshort{rbgs} (bright) for distributed \refi}
	\label{fig:dist_ref_mg}
\end{figure}
%
%Figures \ref{fig:dist_grb_mg}, \ref{fig:dist_ref_mg}, \ref{fig:shmem_arm_alp_mg} and \ref{fig:shmem_arm_ref_mg}.
Figures \ref{fig:shmem_arm_alp_mg}-\ref{fig:dist_ref_mg} show the percentage of time spent in the \gls{mg} kernel, split between the \refres kernels (dark bars) the and \gls{rbgs} kernel (bright bars).
The plots show the results for both implementations and for several setups, both shared memory and distributed; for each amount of compute resources (threads/nodes), the plots show the breakdown for each of the 4 coarsening levels (finer to coarser from left to right, respectively).
The percentages refer to the total execution time, and the runtime in a given level does not include coarser levels.
In all plots, \gls{mg} (thus including \gls{rbgs}) emerges as the main kernel of the entire \gls{hpcg} simulation, with the aggregated \gls{mg} execution time spanning from 80\% to 90\% of the total execution time, while \gls{cg} occupies the remaining 10\% to 20\% of the execution time.
In turn, the  aggregated execution of the \gls{rbgs} smoother across all coarsening levels accounts for always more than 50\% of execution time.

Figures \ref{fig:shmem_arm_alp_mg} and  \ref{fig:shmem_arm_ref_mg} show the results for shared memory on \arms.
Both \refres and the \gls{rbgs} kernels show good scalability with many threads, with their percentage of runtime barely varying with respect to the total execution.
In particular, in \alpi the percentage of the two kernels slightly decreases with the increasing number of cores, while in \refi it fluctuates more clearly, probably due to the \gls{numa}-unaware allocation policy.
The results for \intels are qualitatively very similar, and thus not shown.

Instead, Figures \ref{fig:dist_alp_mg} and \ref{fig:dist_ref_mg} show the results for the distributed-memory setup, where \refi spends a smaller percentage of time than \alpi in \refres but a slightly higher one in \gls{rbgs}.
%We attribute this slight advantage to the usage of MPI-3 primitives and of the task-based OpenMP smoother implementation, which smoothes the values in $ z $ asynchronously as soon as the inputs from the network are available.
We attribute this to the multiple synchronization steps the implementation of \gls{rbgs} within \refi, where for each color each node synchronizes with its neighbors.
Instead, \refres is more expensive in \alpi because it is implemented via an \code{mxv}, which requires synchronization (as from \cref{tab:bsp_costs}), while \refi directly accesses the vector data (as from \cref{sec:restriction_refinement}).
For both implementations the percentages of execution time of the kernels remain very close across different numbers of nodes.

This observation, together with the good weak-scaling of \refi in \cref{fig:dist}, shows that the chosen design scales well.
The worse scalability behavior of \alpi of \cref{fig:dist}, in opposition to the good results in \cref{fig:dist_alp_mg}, instead are caused by an inability to pass domain-dependent information to \grb.

\section{Related work} \label{sec:related}
In the literature, multiple efforts are devoted to the optimization of the \gls{hpcg} benchmark for different architectures.
Architectures based on commodity CPUs are still common targets for \gls{hpcg}, and vendors often made large efforts to showcase the capabilities of their architectures and the optimization strategies suitable for this benchmark: this is the case, among others, for x86 \cite{hpcg_intel} and for ARM \cite{arm_hpcg}.
Similarly, \gls{hpcg} implementations for CPU-based distributed architectures receive particular attention \cite{k_supercomp_hpcg,hpcg_tianhe}.

Another body of work optimizes the benchmark for programmable accelerators like Intel Xeon Phi  \cite{hpcg_intel} or NVIDIA accelerators  \cite{hpcg_cuda_dist} at cluster scale.
Accelerated solutions can be effective to achieve high performance, but need thorough optimizations and may exacerbate load-balancing issues.
Indeed, some work also employs hypergraph partitioning strategies \cite{hpcg_dist_hypgraph} to load-balance among heterogeneous nodes.

All of these solutions face the challenge of parallelizing the \gls{gs} smoother, and typically resort to various implementations of \gls{rbgs} together with a coloring algorithm, as this work does.
Nonetheless, they devote large efforts to perform dedicated optimizations in order to exploit the underlying hardware.
For example, \cite{hpcg_tianhe} stresses on the importance of kernels fusion to improve access locality and save on bandwidth, a typical bottleneck for \gls{hpcg} nowadays.
Although fusion is a common compiler technique to achieve this goal, it is hard to perform it in an automatic fashion on complex code bases, especially if written with general-purpose programming languages like those traditionally used in \gls{hpc}.
Here, exploiting domain-specific information like the algebraic properties of operations required by the \grb specification can expose to the framework and the compiler many optimization and parallelization opportunities.
Indeed, recent work \cite{alp_nb} already proposes an extension of the current \alp implementation that fuses kernels corresponding to different \grb operations, showing large benefits in terms of locality and thus performance.

\section{Conclusions and future work} \label{sec:concl}
This work presents a design of the \gls{hpcg} benchmark on top of \grb, discussing the main kernels and their re-design for an effective implementation on top of standard algebraic operations.
The code is available open-source together with \alp\footnote{\url{https://github.com/Algebraic-Programming/ALP}}.
%We replace the original \gls{gs} smoother with the similar \gls{rbgs} smoother, which can be naturally expressed with algebraic operations.
The implementation on shared memory systems outperforms the reference one, bringing further evidence that expressing algorithms in a high-level, algebraic formulation enables a large spectrum of opportunities for automatic optimizations.
Therefore, domain experts may attain both performance and productivity gains by pursuing an algebraic formulation of their problems. %, rather than focusing on manual optimizations.

Instead, the distributed-memory implementation shows fundamental scalability limitations compared to the reference, which leverages domain-specific geometrical information.
For future work, \cref{sec:enhancements_shmem} discusses possible enhancements that would especially benefit shared memory systems, while
\Cref{sec:future_distrib} discusses the limitations for the distributed-memory parallelisation of HPCG, potential mitigations, and changes those would imply for \grb.

\subsection{Improvements for shared memory systems} \label{sec:enhancements_shmem}
Domain-specific information can be beneficial for performance also in the case of a shared memory system.
For example, the restriction and refinement operations in \grb in \cref{sec:restriction_design} must be materialized into a dedicated matrix, with storage and bandwidth costs.
%For benchmarks in well-known domains like \gls{hpcg}, these operations can be more concisely described with a few parameters, which is how the implementations evaluated here generate the related matrices.
Instead, a \grb framework may be extended to allow a more abstract description of a linear operation, which may enable many compiler optimization and trade bandwidth for computation.
Such an abstraction is possible when an operation can be described with few parameters, which is the case for restriction/refinement if we take into account the geometrical structure of the \gls{hpcg} input.
Although such abstract representation may not be feasible in domains like \glspl{amr}, several workloads that operate on regularly structured sparse matrices, whether geometric in nature or not, may benefit.

Orthogonally, existing abstractions already allow pipelined execution of \grb operations \cite{alp_nb}, with benefits for data locality and performance under investigation.
% ALBERTO_PLEASECHECK: what you describe are what Daniele for example calls access tensors (or matrices). Those are in turn a specific class of a banded matrix structure that we introduced with ALP/Dense (not yet published, so I don't want to refer to it in this paper-- maybe a later version can cite a preprint, if Daniele and co are in time for it). Core point is that it applies to any workload where we have a low-rank structure or a fixed sparsity pattern that fits the union-of-bands class.

\subsection{Future work for distributed systems} \label{sec:future_distrib}
Distributed systems fundamentally challenge the \grb specification, in particular the opaqueness of containers:
when domain-specific, geometrical information is lost or hidden in the nonzero pattern of a matrix, standard matrix distribution strategies usually cannot achieve the same scalability as domain-specific solutions.
We describe four classes of solutions, dubbed i to iv, ordered by how much expertise is required by the \grb user.
Solution A may hence be appropriate for \emph{hero programmers} \cite{humble}, which includes expert \gls{hpc} programmers who typically compose and maintain code like \gls{hpcg}.
On the other end of the spectrum, solution D would be preferred by \emph{humble programmers} \cite{humble} as it allows them to focus on the purely algebraic specification of the code,
and not on details that affect performance or scalability.

\begin{enumerate}[label=\roman*)]
\item A definitive solution is to pass such domain-specific information to a \grb container so to explicitly give the implementation a preferred data \emph{partitioning},
from which \grb may automatically derive a parallel work schedule for level-2 and 3 BLAS operations, both sparse and dense.
Such annotation, however, 1) requires strong programmer expertise not only on their domain but also on the mapping of domain knowledge to linear algebra concepts, while 2) a framework implementing such a solution breaks the current \grb prescription of opaque containers.
We dub this solution A, and proceed to discuss alternatives that alleviate these disadvantages.

\item Limiting the amount of programming effort required, multiple distributed backends with different matrix partitioning strategies may be provided.
For example, a 2D matrix distribution instead of \alp current 1D decreases the communication cost in \cref{tab:bsp_costs} from $ n (p - 1) / p $ to $ n / p \left( \sqrt{p} - 1 \right) $, thus only partially alleviating the communication bottleneck visible in the weak scaling experiment of \cref{sec:exp_dist}.
This only partially improves on the current state and still requires some expert knowledge to understand how efficiently the input domain maps on the backend partitioning strategy.

\item One can leverage existing distributed data structures to describe grids and implement a \grb wrapper for these objects, allowing their use with linear algebraic operations.
However, this solution depends on external software components and thus looses the performance guarantees of \grb.
Another option is to extend \grb with the notion of grids, and provide matrix views of such objects; however, this escapes the intended scope of the \grb specification.

\item Finally, black-box partitioning methods may be used to automatically infer a partitioning of \grb matrices, with little to no user intervention.
This reconstructs the domain information from the input itself, although at potentially significant pre-processing cost incurred while creating a \grb matrix, or may, at the cost of generality, incur this cost at compile time \cite{sympiler}.
\end{enumerate}

% \bibliography{IEEEabrv,biblio}
\balance
\bibliography{biblio}

\end{document}